\pdfoutput=1
\documentclass[fleqn,usenatbib]{mnras}

\usepackage{newtxtext,newtxmath}

\usepackage[T1]{fontenc}
\usepackage{ae,aecompl}

\usepackage{graphicx} 
\usepackage{amsmath} 
\usepackage{amssymb} 

\newcommand{\kms}{km\,s$^{-1}$}

\newcommand{\beii}{Be\,{\sc ii}}
\newcommand{\bevii}{$^{7}$Be}
\newcommand{\beviiii}{$^{7}$Be\,{\sc ii}}
\newcommand{\beviiiii}{$^{7}$Be\,{\sc iii}}
\newcommand{\lii}{Li\,{\sc i}}

\newcommand{\livii}{$^{7}$Li}

\newcommand{\hei}{He\,{\sc i}}
\newcommand{\he}{He\,{\sc}}
\newcommand{\heii}{He\,{\sc ii}}

\newcommand{\niii}{N\,{\sc iii}}
\newcommand{\oiii}{O\,{\sc iii}}
\newcommand{\ovi}{O\,{\sc vi}}
\newcommand{\mgii}{Mg\,{\sc ii}}
\newcommand{\feii}{Fe\,{\sc ii}}
\newcommand{\tiii}{Ti\,{\sc ii}}
\newcommand{\crii}{Cr\,{\sc ii}}
\newcommand{\vii}{V\,{\sc ii}}

\newcommand{\ciii}{C\,{\sc iii}}
\newcommand{\civ}{C\,{\sc iv}}

\newcommand{\nv}{N\,{\sc v}}
\newcommand{\aliii}{Al\,{\sc iii}}
\newcommand{\neiii}{Ne\,{\sc iii}}
\newcommand{\nev}{Ne\,{\sc v}}

\title[  \bevii\ in  V838 Her ]{Absorption and emission features of \beviiii\ in the outburst spectra of
 V838 Her (Nova Her 1991)}
\author[]{ Selvelli, P. $^{1}$, Molaro, P.$^{1}$, Izzo,L.$^2$
\thanks{E-mail: selvelli@oats.inaf.it (PM)},
  \thanks{ Based on INES data from the International Ultraviolet Explorer  Satellite. }\\
$^{1}$  INAF-Osservatorio Astronomico di Trieste, Via G.B. Tiepolo 11, I-34143 Trieste, Italy\\ 
 $^2$ Instituto de Astrofisica de Andalucia, Glorieta de la Astronomia s/n, 18008, Granada, Spain}
\hypersetup{draft}
\begin{document}

\date{Accepted.... Received 2018...}

 \pubyear{2002}

\maketitle
\label{firstpage}

\begin{abstract}
High and low resolution {\sl IUE}  spectra of V838 Her  in
the early outburst stages  exhibit a strong absorption feature shortward of
$\lambda$3130. We discuss the nature of this spectral feature and provide
convincing evidence  that it  corresponds to  the blue-shifted resonance
doublet of singly-ionized  \bevii\,  recently discovered in other novae.
During the evolution of the outburst the appearance of an emission feature
 close to $\lambda$3130 \AA~ is also identified as \beviiii~ $\lambda$3132
because the
usual identification as the \oiii~ $\lambda$3133.7 Bowen fluorescence line is
hardly compatible with both the known  oxygen under-abundance in the nova
ejecta and the low optical depths in the nebula due to the high outflow
velocity.  The average \bevii~
abundance relative to hydrogen, estimated by four different methods, i.e.
the \beviiii/\mgii~ absorption  ratio, and the  \beviiii/\mgii,
\beviiii/\heii$_{1640}$, and \beviiii/H$\beta$  emission ratios is N(Be)/N(H)
$\approx$ 2.5 $\times 10^{-5}$ (by number), i.e. $\approx$ 1.7 $\times
10^{-4}$ by mass. This corresponds to an overproduction of \bevii~ by about
1 dex in comparison
with the  theoretical
models of massive CO  and ONe novae.  Since \bevii\ all converts into
\livii,  the  \bevii/H~ abundance implies a \livii/H overabundance
of about 4  dex over the  \livii/H  meteoritic value  and indicates  a
total ejected mass of  \livii~  of $\approx$ 9.5$\times$10$^{-10}$
$M_{\odot}$. These data are in
line  with  previous observations and indicate  that large amounts of
\livii~  can be synthesized in a variety of novae, including very fast
  ONe novae.

\end{abstract}


\section{Introduction}

Recent studies  of the  spectra of  four novae in outburst  attributed the presence of narrow and
 wide absorption features observed shortward of  $\lambda$3130 to the resonance doublet of \beviiii~
\citep{Tajitsu2015,Tajitsu2016,Molaro2016,Izzo2018}. These  findings
prompted us to inspect the {\sl IUE} archives for the  presence of a
similar strong absorption feature close to $\lambda$3130 in the post
outburst spectra of  novae observed with the International Ultraviolet Explorer
({\sl IUE}, \citet{Boggess1978}).
For this purpose, we  examined all low and high resolution {\sl IUE}
spectra of more than 30 novae in outburst secured in the {\sl IUE}
archive. The low resolution mode of {\sl IUE} (R $\approx$ 5.0 \AA) is
adequate for studying the presence of the  wide absorption component while
the high resolution (R $\approx$ 0.2 \AA) is  adequate to detect the possible presence of narrow absorption
components. Here we discuss the notable case of V838 Her, while a detailed report on the entire dataset will be
presented in a forthcoming paper (Selvelli et al., in preparation).

\section {V838 Her}

V838 Her  was discovered on  March 24, 1991, and faded by 3 magnitudes in
the first five days ($t_3 = 5$ days, \citealt{Vanlandingham1996}), one of
the fastest decline on record.  The UV maximum occurred close to March 31,
seven days after optical maximum \citep{Ingram1992,Starrfield1992,Leibowitz1993}.
 The nova outburst reached a  large expansion velocity, that was greater  than
3000 \kms. \citet{Matheson1993}  and \citet{Vanlandingham1996} on the  basis of the
characteristics of the outburst and of  the observed abundances  suggested
that the outburst occurred on a massive (1.35 M$\odot$) ONe white dwarf
(WD). A  massive  WD is also predicted by \citet{Kato2009} from modelling of
the optical and UV outburst lightcurve. However, \citet{Szkody1994} pointed
out the observational difficulties of deriving a good value for the mass of the primary and from H${\alpha}$
radial velocity solution they found only a lower limit of M$_1$ $>$ 0.62 M$\odot$.
ROSAT PSPC (Position Sensitive Proportional Counter)
observations in the 0.6-1.3 keV range  detected V838 Her on 1991 March 30
(day 6), making it the first nova to be observed in X-ray within 5
days of an outburst \citep{Trumper1990}.
The X-ray emission was
interpreted as shock heating by the  interaction of the nova ejecta with
pre-existing material, although the origin of this material was not clear
\citep{Lloyd1992,OBrien1992}.
{\sl IUE} studies mostly dealt with detailed abundances analysis from
spectra taken during the nebular phase, about one month after the outburst
and later: a notable spectroscopic feature was the significant
under-abundance of oxygen and the overabundance of neon
\citep{Vanlandingham1996,Schwarz2007}.
\section{IUE spectra and data reduction}

The {\sl IUE} databank contains 43 low resolution and 4  high
resolution spectra of V838 Her, obtained  with the short wavelength prime (SWP)
 and the  long wavelength prime  (LWP) cameras.

Table 1 provides  the log of the subset of {\sl IUE} spectra
used in this study (we disregarded the spectra secured in the
transition stage, between day 5 and 10, and those of  the late stages).
They were retrieved from the {\sl IUE} Newly Extracted Spectra (INES) final
archive \footnote{http://sdc.cab.inta-csic.es/ines/}. The more relevant
modifications in the INES system,  in comparison with  the New Spectral
Image Processing System (NEWSIPS) format of the {\sl IUE} final archive
are:
1) the adoption of a new noise model; 2) a more accurate representation of
the spatial profile of the spectrum; 3) a more reliable determination of
the background.  For a detailed description of the {\sl IUE}-INES system
see \citet{RodriguezPascual1999} and \citet{GonzalezRiestra2001}.

Some {\sl IUE} SWP and LWP low resolution spectra were obtained using
both the large and the small aperture in the same image. We
carefully checked all spectra for   absence of saturation effects
in the
continuum or in the emission lines, a phenomenon that may occur longward
 of 1800
\AA~ in  the SWP camera, or close to 2800 \AA~  in the  LWP camera,  where
the sensitivity is higher. In some cases the best exposure was associated
to the small aperture that  suffers of throughput loss. To compare
spectra taken with different apertures, the continuum of small aperture
spectra was scaled  to match the continuum level of the large aperture
spectrum. This is justified because the time separation between the two
spectra is of the order of a few tens of minutes. After these preliminary
operations, good spectra taken in the same date were averaged\
to improve their S/N ratio and SWP and LWP spectra were\
merged.

\begin{table}\label{tab:1}
\caption{Journal of the {\sl IUE}  spectra of V838 Her  utilized in
the
present  work.}
\begin{center}
\scriptsize\begin{tabular}{lrrrlrrr}
\hline
\hline
\multicolumn{1}{c}{Day} &
\multicolumn{1}{c}{{Image}} &
\multicolumn{1}{c}{{Disp.}} &
\multicolumn{1}{c}{{Apert.}} &
\multicolumn{1}{c}{Date}&
\multicolumn{1}{c}{ExpTime} &   \\
\hline
\hline
1  &  LWP19986  &  LOW  &   SMALL &   1991-03-25  &    19.538     \\
1  &  SWP41180  &  LOW  &   LARGE &   1991-03-25  &    299.697    \\
1  &  SWP41181  &  LOW  &   LARGE &   1991-03-25  &    1199.588   \\
3  &  LWP19992  &  LOW  &   LARGE &   1991-03-26  &    29.778     \\
3  &  LWP19992  &  LOW  &   SMALL &   1991-03-26  &    119.480    \\
3  &  LWP19993  &  HIGH &   LARGE &   1991-03-27  &    1799.659   \\
3  &  SWP41192  &  LOW  &   LARGE &   1991-03-27  &    1199.588   \\
3  &  LWP19994  &  HIGH &   LARGE &   1991-03-27  &    3299.615   \\
4  &  SWP41205  &  LOW  &   LARGE &   1991-03-28  &    419.710    \\
4  &  LWP20001  &  LOW  &   LARGE &   1991-03-28  &    39.608    \\
4  &  LWP20001  &  LOW  &   SMALL &   1991-03-28  &    179.691    \\
4  &  SWP41206  &  LOW  &   LARGE &   1991-03-28  &    209.585   \\
4  &  LWP20002  &  LOW  &   LARGE &   1991-03-28  &    19.538    \\
12  &  LWP20085  &  LOW  &   LARGE &   1991-04-05  &    89.579      \\
12  &  LWP20085  &  LOW  &   SMALL &   1991-04-05  &    179.691     \\
12  &  SWP41303  &  LOW  &   LARGE &   1991-04-05  &    359.499     \\
12  &  SWP41303  &  LOW  &   SMALL &   1991-04-05  &    599.524    \\
15  &  LWP20103  &  LOW  &   LARGE &   1991-04-08  &    419.717    \\
15  &  SWP41318  &  LOW  &   LARGE &   1991-04-08  &    359.499    \\
15  &  SWP41318  &  LOW  &   SMALL &   1991-04-08  &    539.723    \\
22  &  LWP20151  &  LOW  &   LARGE &   1991-04-15  &    779.755    \\
22  &  SWP41391  &  LOW  &   LARGE &   1991-04-15  &    719.537    \\
29  &  LWP20209  &  LOW  &   LARGE &   1991-04-22  &    1199.595   \\
29  &  SWP41455  &  LOW  &   LARGE &   1991-04-22  &    1199.588   \\

\hline
\end{tabular}
\end{center}

\end{table}

\section{The  \beviiii\ 3132 \AA\ absorption feature in early  spectra  }

The spectral distribution in the spectra taken in March exhibits an
increase of flux towards longer wavelengths. This is due to the strong
absorption caused by a crowding of lines of singly-ionized metals. There
are also several ''pseudo-emission'' features close to $\lambda\lambda$
2640, 2885, 2980, which in most cases correspond to regions devoid of
strong absorption lines.  The spectral appearance of  V838 Her is  far
more smoothed than in other novae owing to the  high expansion velocity, $\ge$ 3000 km s$^{-1}$. As a
consequence, even the high resolution spectra are of little help in the direct identification of blended
components.

\begin{figure*}
\includegraphics[width=18cm]{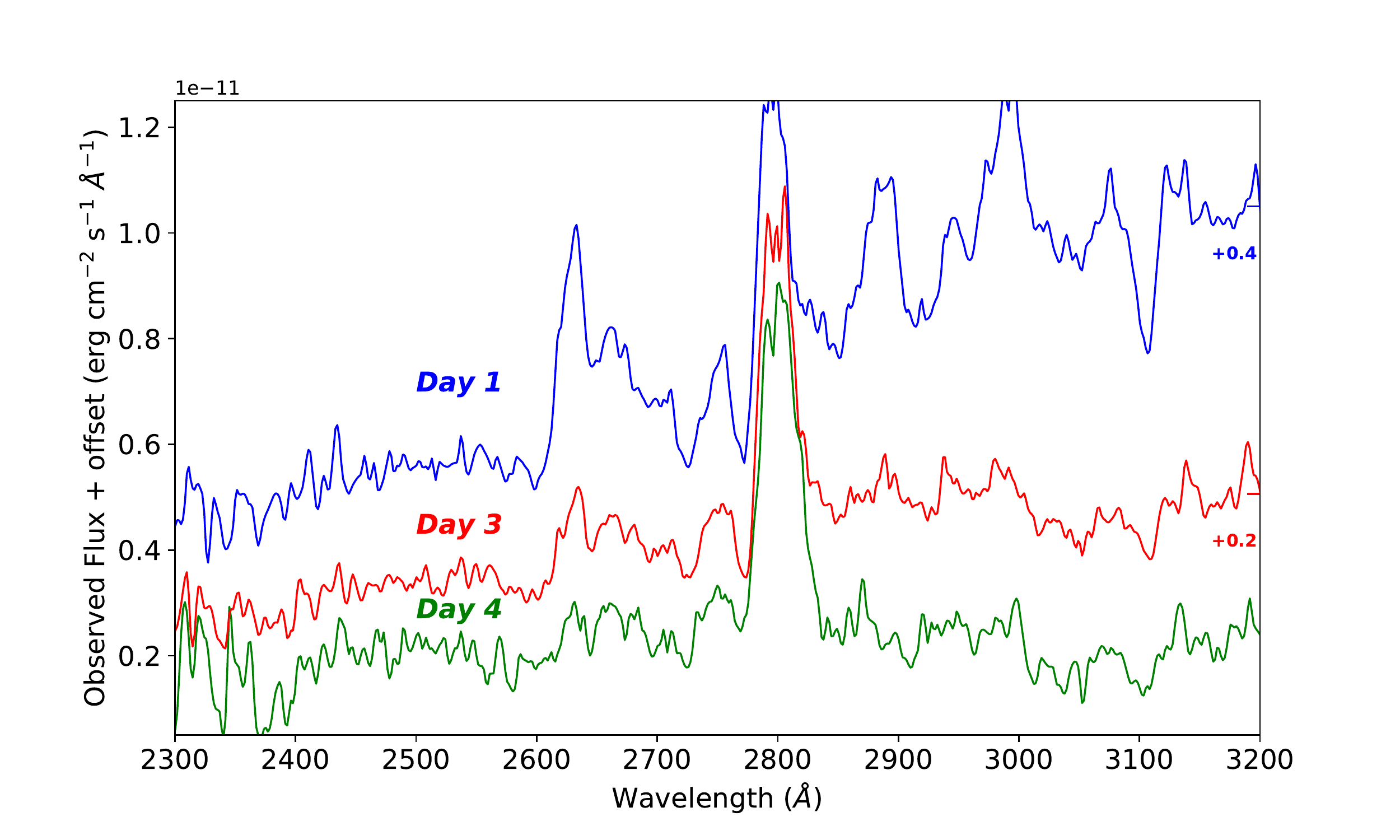}
\caption{The low resolution spectra of V838 Her in the early
absorption stage which exhibit a deep absorption  feature close to $\lambda$3120.
The only definite emission line is \mgii~$\lambda$2800; the other ''pseudo-emissions'' are, probably, windows
 of the continuum.}
\label{fig:1}
\end{figure*}

In previous studies, little attention was paid  to the early phases. Incidentally,
we  note here  that, besides  the resonance doublet of  \mgii~ $\lambda$2796.35 and $\lambda$2803.53,
hereinafter  \mgii~ $\lambda$2800, which
is in emission even in the earliest spectra, some apparent emission lines
close to $\lambda\lambda$ 1240, 1405, 1483, 1550, 1750, 1815, 1900, usually
interpreted as ''windows'' in the iron curtain, will actually correspond to
common emission lines in later spectra.  Remarkable features are the great
flux excess close to $\lambda$1910 \AA, a feature that becomes even more
prominent after correction for the reddening, and the great strength of the
steady absorption close to $\lambda$1850 \AA, identified as the displaced
\aliii~$\lambda$1860 resonance doublet.

Early spectra in both low and high resolution mode reveal the outstanding
presence of a wide, strong absorption feature shortward of $\lambda$3130
\AA~ (hereinafter called $\lambda$3130 line), see
Fig. 1 and Fig. 2.

 \begin{table}
\caption{The absorption equivalent widths of \beviiii~ 3132~   and \mgii~ 2800  
and their ratio in  the early outburst stage of  March 1991.\label{tab}}
\begin{center}
\begin{tabular}{lrrrlrrr}
\hline
\hline
\multicolumn{1}{c}{Date} &
\multicolumn{1}{c}{{Resol.}} &
\multicolumn{1}{c}{{W(\AA)-3132}} &
\multicolumn{1}{c}{{W(\AA)2800}} &
\multicolumn{1}{c}{{3132/2800}} &   \\
\hline
\hline

Day 1  &  Low     & 11.2 $\pm$ 2.0  & 14.6 $\pm$ 0.8   & 0.76   $\pm$  0.14  \\
Day 3  &  Hi      & 10.5 $\pm$ 0.5  & 7.0 $\pm$ 0.3    & 1.50   $\pm$  0.10  \\
Day 3  &  Low     & 11.0 $\pm$ 3.0  & 9.6 $\pm$ 0.7    & 1.15   $\pm$  0.32  \\
Day 4  &  Low     & 11.2 $\pm$ 2.0  & 4.6 $\pm$ 0.8    & 2.43   $\pm$  0.61  \\

\hline
\end{tabular}
\end{center}

\end{table}
We carefully looked for an identification of this absorption feature.

Considering the fact that the relevant spectra correspond to the
''iron curtain'' stage of the nova outburst, the most probable candidates
are the lines of singly-ionized and, in some cases, neutral elements of the
iron group, i.e. \feii, \crii, \tiii, \vii, see Table 3.  However, the interpretation
of the $\lambda$3130 absorption feature as a blend  of absorption lines of only  these species is questionable
for the following considerations:

\begin{itemize}
\item{The absorption strength of the $\lambda$3130 line is comparable to that
of the \mgii~$\lambda$2800 doublet and is stronger than any other absorption
feature. A careful examination of the possible contributors close to
$\lambda$3130, based on \citet{Moore1959,Moore1962} tables and on
the lists of atomic lines on the web sites of
NIST\footnote{http://physics.nist.gov/cgi-bin/AtData/lines-form},
UK\footnote{http://www.pa.uky.edu/~peter/newpage/},  and
 Kurucz's\footnote{https://www.cfa.harvard.edu/amp/ampdata/kurucz23/sekur.
html}, confirmed  that the sole candidates are the lines of
singly-ionized metals of \feii, \crii, \tiii~ and possibly \vii, see
Table \ref{tab:x}. However, lines of these ions arising from the same (or lower) term and with similar or
 higher oscillator strengths do not produce noticeable features in other spectral regions of the nova. We just
mention the case of the many strong lines of \crii~ arising from $\chi$ = 2.5 eV. They may contribute to the
observed $\lambda$3130 feature with about five strong lines, but a similar
or stronger contribution is expected  close to $\lambda\lambda$  2840-2860
taking into account the shortward displacement,  where many strong \crii~
lines with similar lower potential level are present, but only a weak
absorption feature is detectable.
The same reasoning is valid for the lines of \feii: they  contribute  to
the $\lambda$3130 line with only a few lines  of moderate intensity. A
far stronger contribution is expected close to $\lambda$2645 and
in particular
close to $\lambda$2730 where  many strong \feii~ lines are present, but
the observed absorptions in the nova spectra at these wavelengths are
similar or weaker than that close to $\lambda$3130. Therefore, the main contribution to the 3130 feature in V838
 Her does not
originate from  a blend of \crii~ and \feii~ lines.}

\begin{table}
\caption{A list of possible  contributors by single-ionized ions to
the  $\lambda$3130~ absorption feature.}\label{tab:x}
\begin{center}
\begin{tabular}{crrrlrrr}
\hline
\hline
Wavelength & ion & A$_{ij}$ & Level energy\\
\AA &  & $(s^{-1})$ & $(eV)$ \\
\hline
\hline
   3111.597 & TiII  &    2.667e+07   &    1.231332   \\
   3112.953 & TiII  &    2.327e+07   &    1.224145   \\
   3115.200 & FeII  &    6.384e+06   &    3.888695   \\
   3117.483 & FeII  &    6.990e+06   &    3.891610   \\
   3118.580 & TiII  &    5.937e+07   &    1.242992   \\
   3119.550 & CrII  &    1.713e+08   &    2.421358   \\
   3121.264 & CrII  &    1.504e+08   &    2.434119   \\
   3125.879 & CrII  &    8.186e+07   &    2.454790   \\
   3129.599 & CrII  &    8.149e+07   &    2.434119   \\
   3132.961 & CrII  &    8.151e+07   &    2.482828   \\
   3133.958 & FeII  &    1.599e+06   &    3.888695   \\
\hline
\end{tabular}
\end{center}
\end{table}

\begin{figure*}\label{fig:2}
\includegraphics[width=1.9\columnwidth]{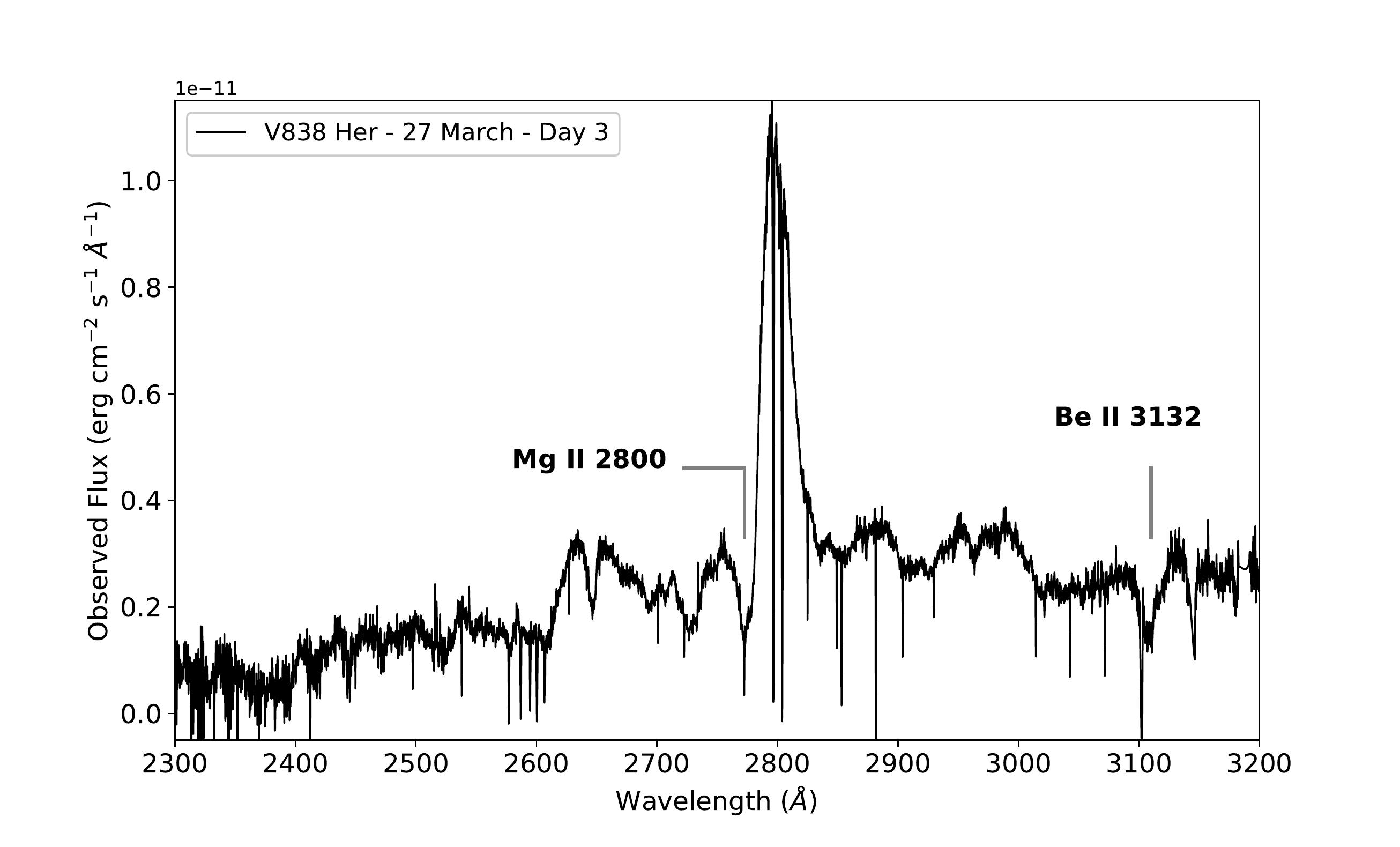}
\caption{The high resolution spectrum of day 3, which exhibit  wide
$\lambda$
3132 and $\lambda$2800 absorption features and  sharp \mgii~ interstellar
 lines, not resolved in Fig. \ref{fig:1}. The other wide absorption
features are
blends of hundreds of strong \feii~ and \crii~ lines, broadened by the high
expansion velocity. The plotted spectrum is the average of the two spectra
LWP19993 and LWP19994 taken in sequence.}
\end{figure*}

\item{
In the early stages the  spectra of  novae  are those  of an optically
thick, expanding,   cooling shell  and resemble those of early type
stars, with blue-shifted absorption lines \citep{Warner2008,Schwarz2001}.
A  visual inspection  of the  {\sl IUE}  low resolution spectra of A-shell
stars has confirmed this and shown a general  similarity  in their overall
line features and spectral distribution with  the spectra of novae during the early ''iron curtain'' stage. 
Besides the strong absorption features close to $\lambda\lambda$  2350, 2580, 2660,  2750, and 2870 which
are all shortward displaced in novae,  they  generally  exhibit also the same
pseudo-emission features close to $\lambda\lambda$ 2640, 2900, 3000,
usually observed also in novae at shorter wavelengths, and commonly interpreted
as consequence of the limited number of absorption lines in these regions,
see also Fig. 3.
These spectroscopic similarities indicate similar ionization and excitation
conditions. In novae, however, the strong line broadening due to expansion
prevents a direct identification of the component absorption lines except
by spectral modeling. Instead, the  limited line broadening of these
A-shell stars and the adequate resolution of the high resolution {\sl IUE}
spectra allow a detailed identification of the relevant absorption
features. Therefore, one can use the {\sl IUE} high and low resolution
spectra
of A-shell stars as a template to interpret the nature  of the\
absorption features observed in novae during the iron curtain phase. For
this purpose, we studied the UV  behavior close to $\lambda$3130 and in
nearby regions of a wide sample of 35 A-shell stars.

Our analysis  indicates no evidence of strong absorption features close to
$\lambda$3130, where the overall aspect is similar to that in adjacent
spectral regions without any concentration of lines in this range.  The
absorption lines close to $\lambda$3130 are easily identified  as transitions of \crii~,
opt.(5), and
\feii, with minor  contributions by \tiii~  and  \vii~, mainly from
the
 complete Kurucz's database.
For all 35 objects the comparison  between the absorption strength in a 30
\AA~ wide region centered close to $\lambda$3130  with the absorptions
close to
 $\lambda$2750 where many strong  \feii~  lines are present  and
$\lambda$2860 where many absorption lines of  \crii~  from the same lower
level as the lines close to $\lambda$3130 are present, clearly
shows  that the absorption lines close to $\lambda$2750  and
$\lambda$2860
 are systematically more numerous and stronger than those close to
$\lambda$3130, and produce deeper absorptions than in the 3130 region.
Instead, the contrary is observed in the spectrum of V838 Her:  the
$\lambda$3130 absorption is  far stronger than the other two.
In A-shell stars we also observe a crowding of absorption lines close to
$\lambda\lambda$ 2620-2640, most of which are identified as \feii~ and \crii.
However the same objects exhibit only weak absorptions close to
$\lambda$3130, see Fig. \ref{fig:3}. On the contrary, in V838 Her the
$\lambda$3130 absorption
feature is stronger than that close to $\lambda$2665.

These findings are  a clear proof of the
presence of an
additional strong absorption  contribution close to $\lambda$3130, unless a
peculiar and  unidentified spectroscopic mechanism acted to reinforce the
absorption lines of \crii~ and \feii~  close to $\lambda$3130 only in V838 Her,
without reinforcing the other absorption lines of the
same ions arising from the same or even lower terms. }

\begin{figure*}
\includegraphics[width=1.9\columnwidth]{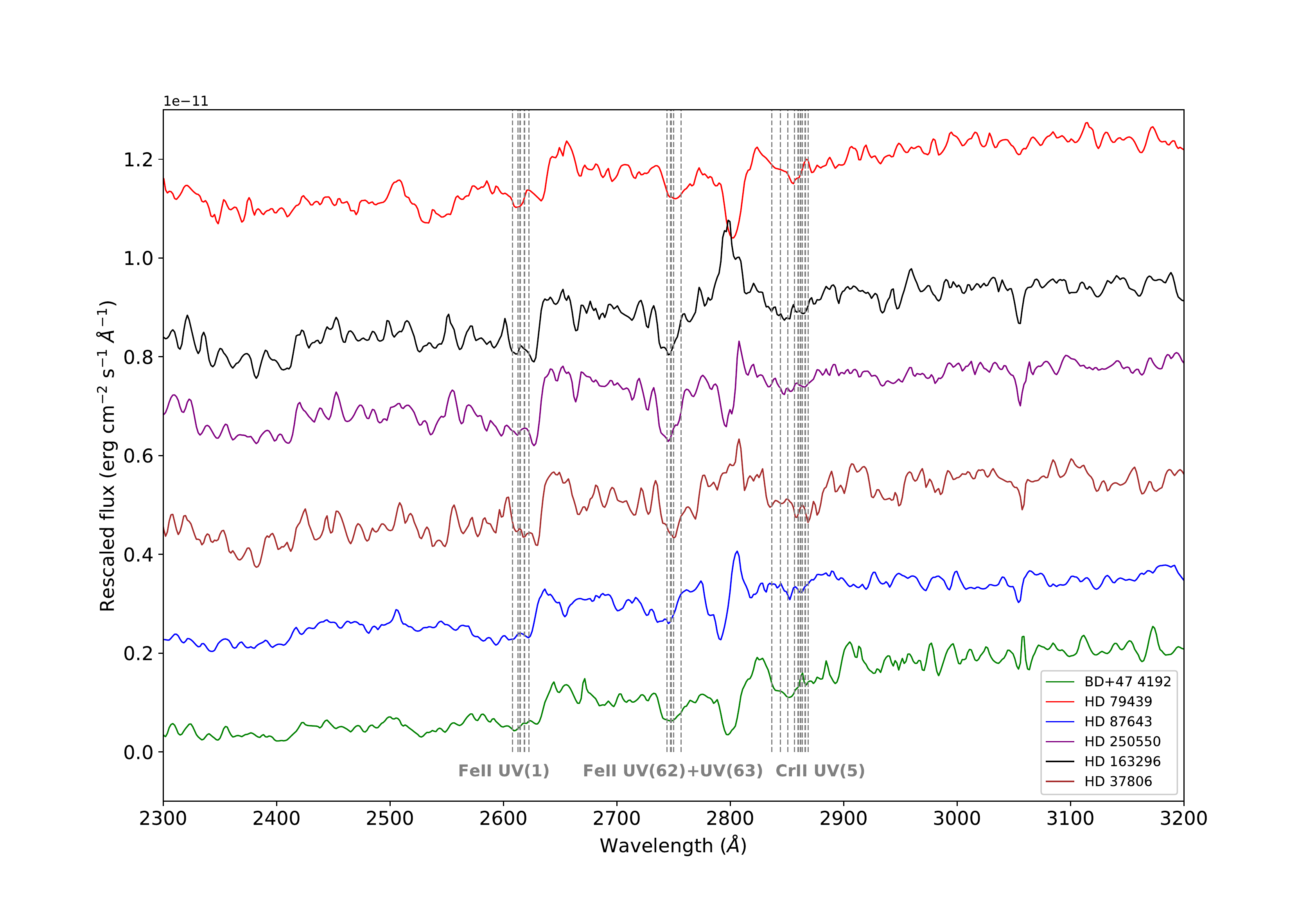}
\caption{IUE low resolution spectra of six A-shell stars in the near UV. The
 $\lambda$3130 absorption
feature is  weak or absent while other absorption features, e.g.
$\lambda$2750 and  $\lambda$2850,  are stronger compared to those in V838
Her.}
\label{fig:3}
\end{figure*}

\item{
    The PHOENIX models \citep{Hauschildt1997} correctly reproduce
the absorption features  observed in the early {\sl IUE} spectra of several
novae but fail to reproduce the absorption feature close to $\lambda$3130,
which is present in several spectra (see  Figs. 5 and 6 in
\citet{Hauschildt1994}; Figs. 5, 7, and  10, in \citet{Schwarz1997}, Fig. 6
in \citet{Schwarz1998}; and Figs. 7, 8 and 10 in \citet{Schwarz2001}
}

\item{
In the high resolution spectra taken in the early phase, despite of the
uncertainty associated with the correction  of the echelle orders close to
$\lambda$3130,
there is a definite  similarity between the profiles of the absorption
features shortward of \mgii~$\lambda$2800 and $\lambda$3130. If we refer
to the nominal wavelengths of \mgii~$\lambda$2800 and \beviiii~$\lambda$3132,
the lines exhibit a similar velocity field structure and similar velocity
displacement close to ~-3000 \kms, see Fig. 4. We consider the similarity
in the profiles as another clear indication of the  same spectroscopic
nature of the two features, that is,  a resonance transition.
Needless to say, there are no resonance lines close to  the $\lambda$3130
region except
the \beviiii~ doublet $\lambda$3132.
}

\end{itemize}
\begin{figure*}
\includegraphics[width=1.9\columnwidth]{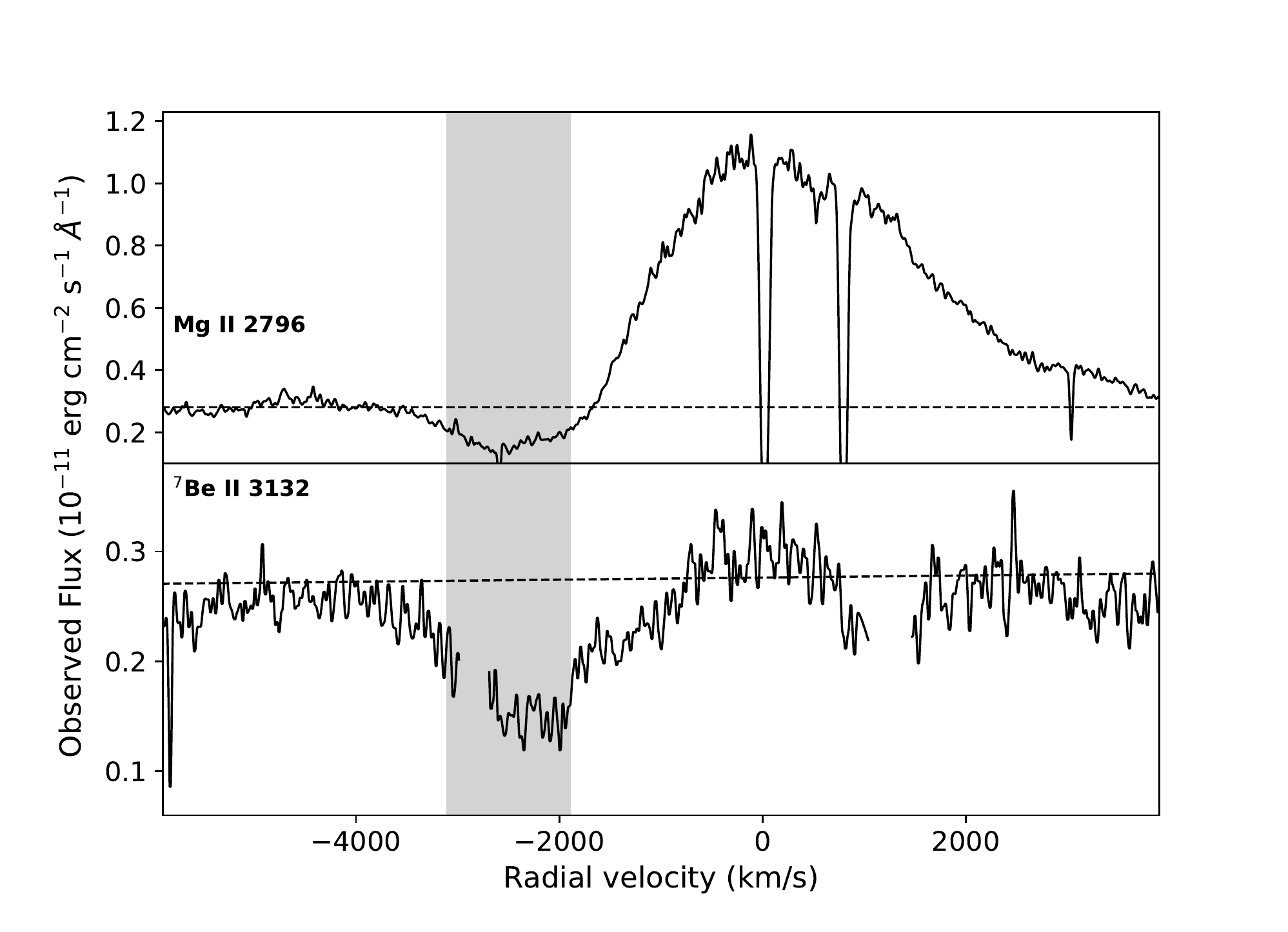}
\caption{The P-Cygni absorptions of  the \mgii~$\lambda$2800 and
\beviiii~$\lambda$3132 lines in the {\sl IUE} high resolution spectrum  of day 3.
Note the similar velocity structure of the two resonant
transitions.  The gap close to -3000 km s$^{-1}$  in the spectrum of \beviiii  corresponds
 to the absence of overlap between two adiacent echelle orders.
}
\label{fig:Fig-4.png_figure}
\end{figure*}

In conclusion, atomic data and observations indicate that the strong and
wide absorption feature close to $\lambda$ 3130 in V838 Her can be
only
partially explained as a blend of common iron curtain absorption  lines.
This  agrees with the studies of recent novae by
\citet{Tajitsu2015,Tajitsu2016} and \citet{Molaro2016}, who found only a
minor contribution of singly-ionized metal transitions to this feature.

A visual comparison between the absorptions close to $\lambda\lambda$ 2660,
2750, 2860 and 3130 in  A-shell stars and in V838 Her indicates that,
on average, only a fraction of the $\lambda$3130 absorption of V838 Her
is produced by  lines of the Fe-peak elements.

Therefore,  we confidently identify the 3130 feature as due  mainly to the
\beii~  resonance doublet and assume that it is \beviiii~  $\lambda$
(vac.)   3131.49 and $\lambda$  (vac.) 3132.13 (hereinafter
\beviiii~$\lambda$3132)    on the basis of this detection, from  narrow
components,  in  other novae.

\section{ The \bevii\  3132 \AA ~ emission line in late spectra}

An emission feature centered close to $\lambda$3130 is observed in the low
resolution spectra taken in April 1991, see  Fig. \ref{fig:5}.

This emission feature is quite commonly observed in planetary nebulae,
symbiotic  stars, novae and other emission-line sources and is usually
identified as \oiii~$\lambda$3133.70, the strongest line produced in
the Bowen fluorescence mechanism (BFM). For a general description of the
BFM
see \citet{Osterbrock2006}  and  \citet{Dopita2003}, while for a detailed
analysis of the BFM in the symbiotic nova RR Tel  see \citet{Selvelli2007}.

However, in the case of V838 Her the identification of the
$\lambda$3130 emission feature as the \oiii~$\lambda$3133.70 fluorescent
line is improbable for these reasons:

\begin{itemize}

\item
{ All UV and optical studies  of V838 Her noted the absence
or extreme weakness of  the \oiii~ lines
\citep{Matheson1993,Starrfield1993,Vanlandingham1996,Schwarz2007}.
\citet{Downes2001}, from  a study of 96 novae, pointed out that V838 Her
 was an outlier in the sample because the
[\oiii]~$\lambda$ 5007~ line
was
about 100 times fainter than all other novae. This can also be seen  from
Fig. 4 of \citet{Williams1994a} and Fig. 1 and 2 of \citet{Iijima2010}.
The inspection of the {\sl IUE} SWP spectra of  V838 Her taken in
April confirms beyond  any doubt the absence of the common
\oiii]~$\lambda$1660-1666 emission lines, while other common
emission lines, such as \civ~$\lambda$1550, \niii]~$\lambda$1750, and
 \ciii]~$\lambda$1906, in a range of ionization that encompasses that of
\oiii, are clearly present (see Fig. 6).

This surprising peculiarity of V838 Her was  pointed out
recently\footnote{stsci.edu/$\sim$ofox/posters2017/posters/starrfield/poster.pdf}.

We  considered the possibility that the absence of the \oiii] and
[\oiii]  emission lines could simply be caused by  a higher-than-critical
electron density in the nebula, a  situation in which the optical forbidden
and the UV intercombination emission lines of  \oiii~ are suppressed  by
collisional de-excitation.   This is unlikely,  because  the
UV
\oiii]~$\lambda$1666 line has $n_e^{crit}$ of the order of $4.6 \times
10^{10}$
el. cm$^{-3}$, a value that is higher than that implied by the presence of
the intercombination lines of \niii~$\lambda$1750  ($2\times 10^{10}$ el.
cm$^{-3}$), and \ciii~$\lambda$1908  ($3.2\times 10^9$ el. cm$^{-3}$),
which are prominent in April spectra. Therefore, the faintness of the
\oiii~ lines is a consequence of  the under-abundance  of oxygen, in
particular
when compared to  the abundance of carbon and nitrogen,  (see Fig. 6)}

\begin{figure*}
\includegraphics[width=18cm]{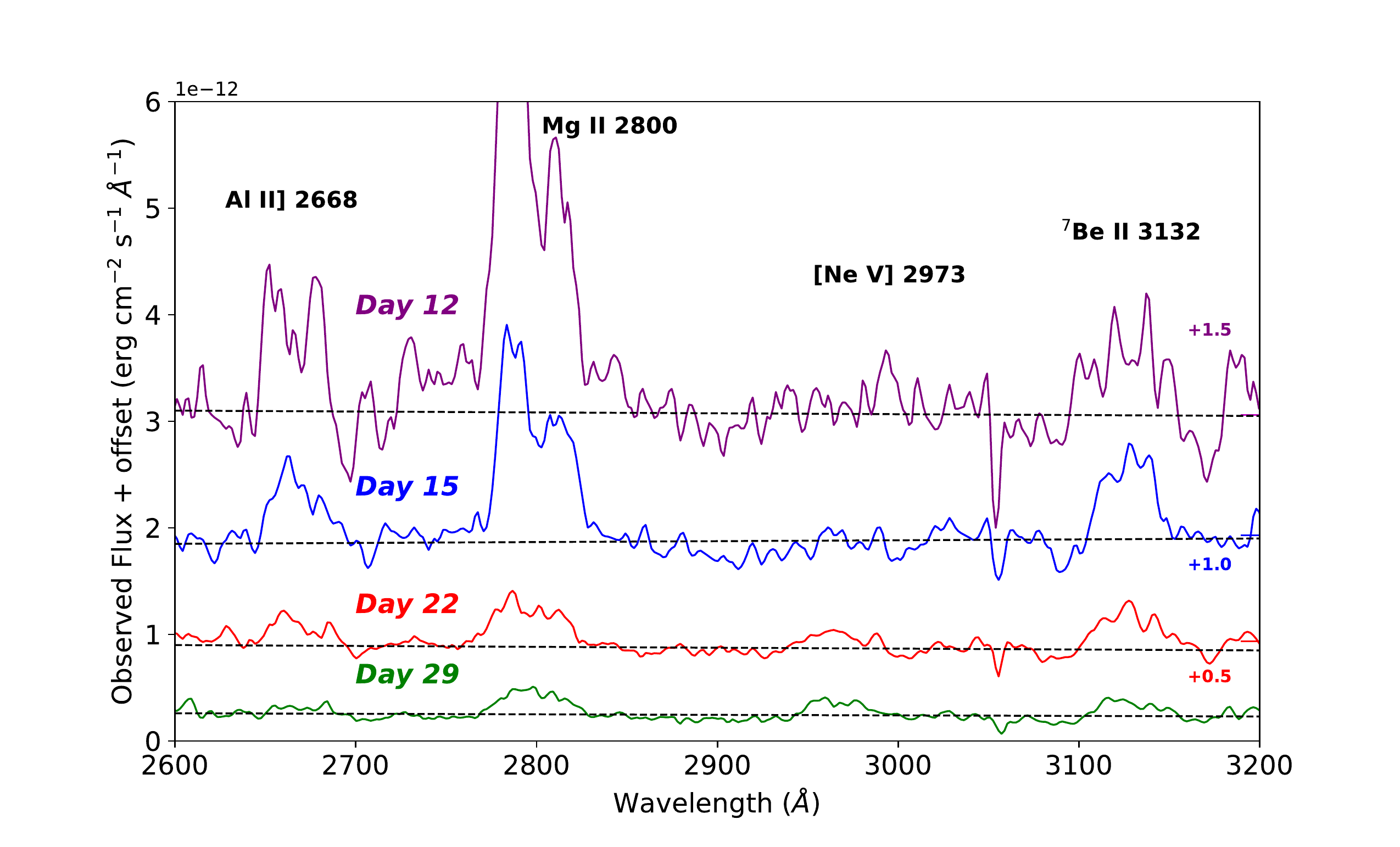}\label{fig:5}
\caption{The \mgii~$\lambda$2800 and \beviiii~$\lambda$3132
emission lines in the low resolution spectra of April 1991. }
\end{figure*}

\begin{table*}
\small{
\caption{The emission line intensities (10$^{-11}$ erg cm$^{-2}$ s$^{-1}$
\AA$^{-1}$)  of  HeII~$\lambda$1640, \mgii~$\lambda$2800 and \beviiii~
 ~$\lambda$3132  and their ratios in April  1991. All measurements are
 from low  resolution spectra.
 The observed intensities  were  corrected for  E(B-V)=0.40.
\label{tab}}
\begin{center}
\begin{tabular}{lrrrlrrr}
\hline
\hline
Day & I-1640 & I-2800 & I-3132 & 3132/2800 & 3132/1640 \\
\hline
\hline
 12 & 22.2 $\pm$ 1.0  & 9.7 $\pm$ 0.2 & 5.5 $\pm$ 1.1 & 0.57 $\pm$0.11 & 0.25 $\pm$0.05 \\
 15 & 21.2 $\pm$ 1.1 & 4.3 $\pm$ 0.2 & 2.8 $\pm$ 0.2 & 0.65 $\pm$0.06 & 0.13 $\pm$0.01  \\
 22 & 8.7 $\pm$ 0.5  & 1.3 $\pm$ 0.1 & 1.5 $\pm$ 0.2 & 1.15 $\pm$0.18 & 0.17 $\pm$0.03  \\
 29 & 5.1 $\pm$ 0.3  & 1.25$\pm$ 0.1 & 0.8 $\pm$ 0.2 & 0.64 $\pm$0.17 & 0.16 $\pm$0.04  \\
\hline
\end{tabular}
\end{center}
}
\end{table*}
\vspace{0.3cm}

\item {
The BFM efficiency $\epsilon$ is defined as the  fraction of the \heii~
Ly$\alpha$ photons which are converted  to \oiii~ photons, and  can be
estimated from the intensity ratio of a Bowen line (generally the
$\lambda$3133.7  line) to that of a \heii~ recombination line, e.g.
$\lambda$1640. In this case  $\epsilon$  = 2.25 $\times$
I$_{3133.7}$/I$_{1640}$.

The data of Table 4, using the  \texttt{numpy}
\citep{Numpy}, and the  \texttt{astropy}
\citep{Astropy2013} packages, and
assuming that
the $\lambda$3132 emission were due to  \oiii~ $\lambda$3133.7, would
 give  a weighted average intensity ratio
of    \oiii~ $\lambda$3133.7 to \heii~   $\lambda$1640 =  0.17 $\pm$ 0.03. 
This would give   an
average  conversion efficiency of $\epsilon$ = 0.38 $\pm$ 0.10. 
  This value is close to that
found in planetary nebulae ( see
\citealt{Weymann1969,Likkel1986,Liu1993,Dopita2003}). However, the Bowen
fluorescence process requires low escape probability, i.e. high optical
depths $\tau$, in both the \heii~ and the \oiii~$\lambda$304 resonance
lines, and the efficiency  is related to the optical depths of these
resonance lines. This depends on the  number of absorbers, i.e. nebular
mass and elemental abundances  and on the expanding velocity v$_{exp}$.
PNe have mass ejecta close to 0.1-0.2 M$_{\odot}$ and  low expansion
velocities  of approximately 20 \kms,  and so high $\tau$. On the
contrary,  Nova Her 91  has v$_{exp}$  $\approx$  3000-4000 \kms~ and  a
shell  mass of  $\approx 10^{-5} M_{\odot}$
\citep{Schwarz2007,Vanlandingham1996}.}

\end{itemize}

A full treatment of the BFM was developed only for quasi
static-nebulae where line broadening is mostly by thermal velocities and
detailed non stationary models do not exist \citep{Kallman1980,Dopita2003}.
This would require a detailed  study of the absorption and scattering  of
the relevant resonance photons  in an expanding atmosphere, a quite complex
theoretical task that is usually faced using the Sobolev approximation.
This  is beyond the scope of this paper. We just note  that both
\citet{Weymann1969} and \citet{Kallman1980} pointed out that velocity gradients
in the nebula may decrease the optical depth in the line centres and
increase line escape probabilities, substantially suppressing the
efficiency of the BFM. Macroscopic velocity fields and velocity gradients
increase the escape of photons, substantially suppressing the efficiency
of the BFM if turbulent or flow velocities become comparable  to thermal
velocities,  or higher, (see \citealt{Kallman1980,Peraiah2001,Dopita2003}.

From the observational side,  \citet{Likkel1986} noted
 that the optical depth of the Bowen \oiii~ transitions can be reduced
by differential velocities such as expansion velocity, and pointed out that
PNe with low efficiency have higher expansion velocities. The same
conclusion, from a large sample of PNe, was reached by \citep{Liu1993} who
found that in PNe the efficiency drops abruptly when the expansion exceeds
some 30 \kms.  \citet{Stickland1981},  in a study of Nova Cyg 1978, noted
that for a nova the transfer problem may be complicated due to possible
large velocity gradients, which will tend to reduce the efficiency.

In RR Tel, \citet{Selvelli2007}  found  $\epsilon \approx 0.30$ but here
the lines are sharp and the large number of scatterings in the resonance
lines of \heii~ and \oiii~ favours the conversion of \heii ~$\lambda$303.78
 to \oiii~$\lambda$303.80 and then  to \oiii~$\lambda$3133.7.

In this respect, it is worth noting that long ago \citet{Swings1942}
attributed the absence of BFM in the envelopes of WR stars, where \oiii~
lines were commonly observed,  to their high expansion velocity. In their
words, "the O$^{++}$ atoms located at a specific place in the WR shell are
able to absorb the resonance radiation of only a small fraction of ejected
He$^{+}$ ions, since these must have a definite radial velocity with
respect to the O$^{++}$ atoms considered". \citet{Swings1942} concluded
that in the
case of spherical symmetry  and of large constant expansion velocity BFM
should be practically absent.

We argue that the reported presence of the BFM in some novae, if  the
identification of the 3130 emission feature as \oiii is taken for granted,
 may be explained by the large overabundance of oxygen  in the
ejecta of most novae, by a factor of up to 100-1000 compared to solar. This
 increases the optical depth in the \oiii~$\lambda$304   resonance lines
and compensates the negative  effects of the expansion.
The correlation between high abundances and BFM efficiency was noted
by \citet{Netzer1985} in a study of BFM in AGNs, and explained by the
higher optical depth in the \oiii~ resonance lines. In Nova Cyg 1978
\citet{Stickland1981} found efficiency  close to 0.80   but the ejecta  was
overabundant in oxygen by a factor  25 and  the expansion velocity
was lower (v$_{exp} \sim 1000$ km s$^{-1}$) than in V838 Her.

Based on the two above  mentioned considerations, the absence  of \oiii~
alone being sufficient,  we confidently assume that \beviiii~$\lambda$3132
is the only plausible identification for the emission close to
$\lambda$3130.
Therefore, we conclude that both the shortward-displaced $\lambda$3130
absorption line, observed in the early stages, and the emission  line
centred close to $\lambda$ 3130, observed in the later stage,  are
identified
as the resonance transition of \beviiii.

We thank  the referee for pointing out that the absence of other Bowen lines,
i.e  $\lambda$ 2836  and   $\lambda$ 3047,  may also be evidence for the lack of
 BFM. These lines, although  weaker by a factor of about 7 compared to
 $\lambda$  3133.7,  (see \citealt{KastnerBhatia1996}, and \citealp{Selvelli2007}) are expected to give  a detectable emission if the feature
close  to  $\lambda$ 3132 were the Bowen \oiii~$\lambda$  3133.7 line.

\begin{figure*}\label{fig:6}
\includegraphics[width=18cm]{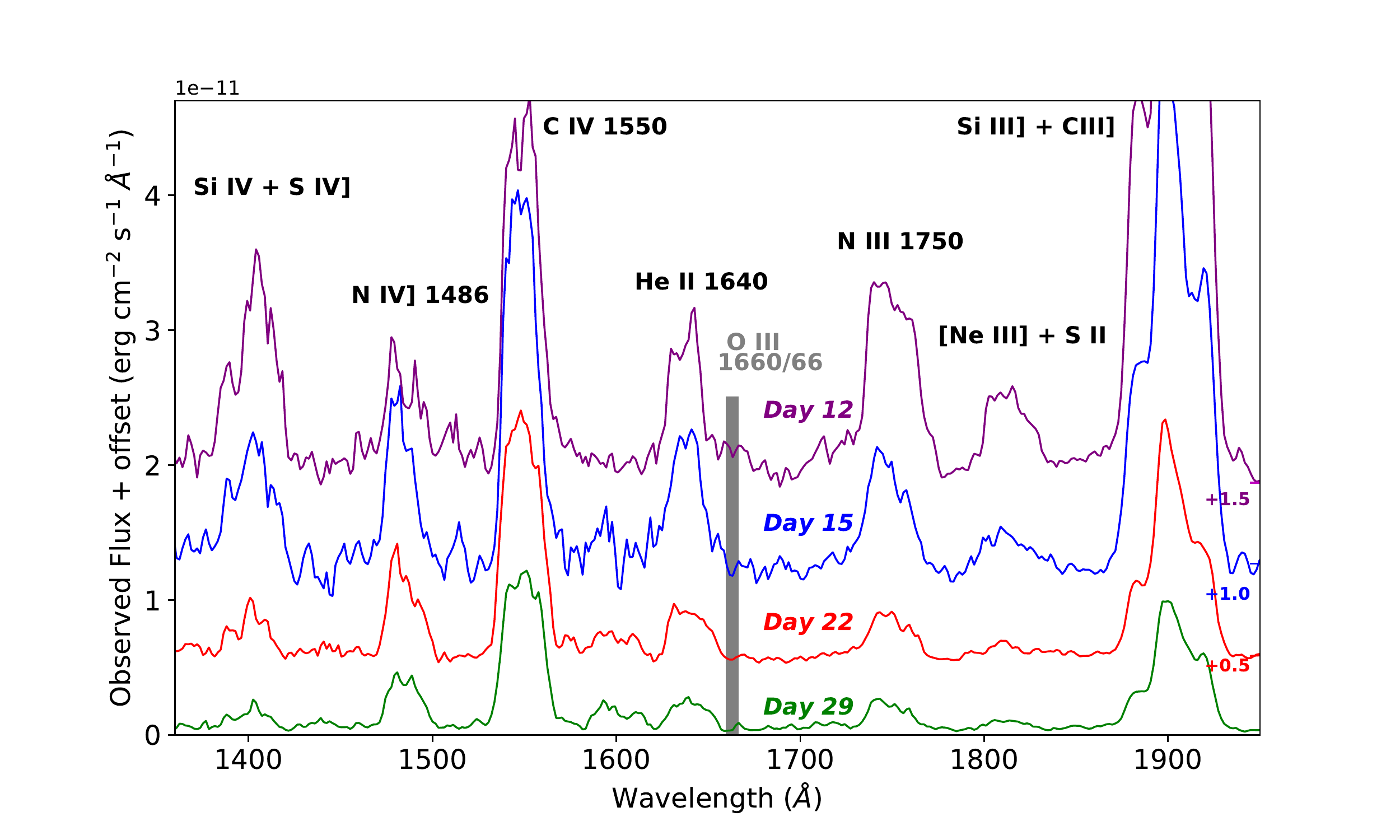}
\caption{The emission lines of the far UV low resolution spectra
in April 1991. The absence of the commonly observed   \oiii]~$\lambda$
1660-1666
emission feature is remarkable.}
\end{figure*}

\section{\bevii\  Abundance  estimate }

We  estimate the  abundance of \bevii\ both by comparing
the equivalent widths of the  \beviiii~$\lambda$3132 with
\mgii~$\lambda$2800  absorption lines in the early  spectra of March,  and
then, in the spectra of April, by comparing the emission intensity of
\beviiii~$\lambda$3132 with those of \mgii~$\lambda$2800,
H${\beta}$~$\lambda$4861, and \heii~$\lambda$1640.

\subsection{ The absorption stage}

Beryllium  and  magnesium  have similar ionization potentials and  a  rough
estimate of the   their relative abundances   can be derived from the ratio
of the equivalent widths (EWs) of the  \beviiii~$\lambda$3132 ($W(3132)$) and
\mgii~$\lambda$2800 ($W(2800)$) absorption doublets.
From the data of Table 2,   using   the \texttt{numpy} \citep{Numpy},
\texttt{matplotlib} \citep{Matplotlib} and the \texttt{astropy}
\citep{Astropy2013} packages  we  derive a ($\sigma$)$^{-1}$  weighted
average EW ratio of 1.80 $\pm$ 0.68, and a ($\sigma$)$^{-2}$ weighted
average EW ratio of 1.28 $\pm$ 0.36,   that we round to 1.5 $\pm$  0.5.

It must be noted  that the  observed  absorption EW  of \mgii\ could be
partially reduced  by  the  presence of the emission  component. Moreover,
as explained by the considerations of Sect. 4, and Table 3,  a minor but  not negligible
fraction of the measured EWs of the $\lambda$3130 line, that we estimate of the order of 10 to 20 percent,  is
also
contributed
by lines of singly-ionized elements e.g \crii, \tiii~ and \feii.
Therefore, the measured ratio of the EWs of \beviiii~$\lambda$ 3132  over
\mgii~$\lambda$ 2800 must be accordingly reduced  by a factor that, following \citet{McLaughlin1941} precept of
"erring
on the side of conservatism",  we   assumed  to have   2.0 as upper limit.
Therefore,  we  adopt  an average corrected ratio $W(3132)/W(2800)$
$\sim$ 0.75 $\pm$ 0.25.

In the case of  optically thin conditions, the equivalent widths can
provide a direct estimate of the number of absorbers ($N_{i}$) through  the common
relation:

\smallskip
\begin{equation}
$W$   \propto   N_i \times f_{ij} \times \lambda^2
\end{equation}

\smallskip
\noindent
Therefore, the relative  population of the ground terms of the two ions
can be easily derived using the relevant values of wavelength and
oscillator strength ($f_{ij}$).

\noindent
Since $f_{ij}$(3132)/$f_{ij}$(2800)$\sim$0.5  and
$W(3132)/W(2800)$=0.75 $\pm$ 0.25, it follows  that
\smallskip
\noindent
\begin{equation}\nonumber
N_i(3132)/N_i(2800) =  1.20 \pm 0.40
\end{equation}
\smallskip
\noindent
If the two ions are mostly in their ground term,  the above  derived  ratio
provides a reliable estimate of  the ratio of ionic abundances. In
addition,
since the second ionization potentials of the two ions have similar
values  (i.e 18.21 eV and 15.04 eV, respectively), similar ionization fractions   are
expected. Thus, the above
derived
ratio
also  provides  an estimate of the total  \bevii/Mg abundance.
The magnesium abundance in  V838 Her is almost solar
\citep{Schwarz2007}, 3.8 $\times$ 10$^{-5}$  by number
\citep{Grevesse1998}, and so   the \bevii~ abundance  in V838 Her relative
to hydrogen, by number,   is given by:
\smallskip
\noindent
\begin{equation}\nonumber
N(^{7}Be)/N(H) = 3.8\times10^{-5}\times1.20(\pm0.40) \sim
4.6\pm1.5\times10^{-5}
\end{equation}

\smallskip
\noindent
A  criticism of this approach could arise from  the fact that the  above
relation   between $W$  and $N_i$  is valid  for optically thin
conditions,  while   optical depth effects could be expected  in  the two
resonance lines. However, the high resolution spectra do not exhibit
indication of saturation  effects in the two lines: the line central depth
is about 0.5 of that of  the continuum level and indicates that the equivalent width  falls
in the linear part of the curve of growth. In general, saturation
effects
are significantly  reduced by the  presence of a strong  outflow velocity and
in V838 Her  v$_{out}$ is greater than 3000 km $s^{-1}$.  Also,
the similarity between two lines which have similar EWs and exhibit a
similar
velocity profiles indicates that the two features are produced under similar
conditions. This  should alleviate the above reported  criticism .

Another source of error arises from the  uncertainty in the determination
of the continuum to which the EWs are measured. However,
from Fig. 2 we see that the  continuum  is clearly  determined  in the
high resolution  spectra  and  at  wavelength close to $\lambda$3000 \AA~
the
iron curtain effects are lower. Also, model atmospheres indicate that close
to
$\lambda$3000 \AA~ the local continuum is similar   to that of  the black
body (see fig. 12 and 13 in \citealt{Hauschildt1997}). In any case, since
the uncertainties regarding \beviiii~$\lambda$3132  and \mgii~$\lambda$2800
are comparable, they compensate each other in the final ratio.

\subsection{The emission stage}

In April  spectra of  the emission lines allow an estimate of the
\bevii~ abundance  from the comparison of  the relative emission intensity
of the  $\lambda$3132 line with those of the   \mgii~$\lambda$2800,
H${\beta}$~$\lambda$4861 and \heii~$\lambda$1640 lines. To use
ratios of emission line intensities  at different wavelengths the spectra
must be de-reddened. Previous determination of the reddening E(B-V) from
the
literature  are in the range  0.50 $\pm$  0.10 mag
\citep{Vanlandingham1996,Kato2009}.

We  re-determined this quantity using the method based on the removal
of the  wide $\lambda$2175 \AA~  interstellar  absorption bump. Using SWP
and LWP spectra taken at the same epoch,   and using the  reddening curve of
\citep{Cardelli1989} we  estimated a colour excess E(B-V)=0.45$\pm$0.05
mag, in agreement  with previous values. IR maps
\citep{Schlafly2011} indicate an upper limit of E(B-V)=0.37  mag for the
total galactic absorption in the direction of V838 Her. Therefore, we
 assumed E(B-V)=0.40 mag as a final value and corrected  the
observed  intensities accordingly.

\subsubsection {The \beviiii/\mgii~ ratio}

The emission lines  of the resonance  transitions of  \mgii~$\lambda$2800
and  \beviiii~$\lambda$3132  can be described by a simple two  level atom
in which the collisional excitation rate is in equilibrium with the rate of  radiative decay
and collisional de-excitation.

\smallskip

\begin{equation}
N_1 N_e q_{12} = N_2 A_{21}  + N_2 N_e q_{21}
\end{equation}

\smallskip
\noindent

Here    $q_{12}$  and $q_{21}$   [cm$^3$ s$^{-1}$]
are the collisional  excitation and de-excitation rate  coefficients,
A$_{21}$ [$s^{-1}$] is the  Einstein coefficient for spontaneous decay  from level 2,
and  N$_e$ [cm$^{-3}$]  is the electron density.

For $ N_2 N_e q_{21}<<N_2 A_{21} $, that is, for  $ N_e<<N_{e}^{crit} =
A_{21}/q_{21} $ one can neglect the second term  in eq. (2)  and
assume that every collisional excitation is followed by a
radiative decay. In this case  the line emissivity is:

\smallskip

\begin{equation}
I_{21} = N_1 N_e q_{12} h\nu_{12}= N_2 A_{21} h\nu_{12} ~      [erg~ cm^{-3} s^{-1}]
\end{equation}

\smallskip
\noindent

For the case of the  resonance  transitions of \beviiii~ and \mgii~,
this is justified by the fact that  their  critical electron
density   $A_{21}/q_{21}$  is  high, i.e.  of approximately of $10^{14}$
el cm$^{-3}$, while  $N_e$ in the nebula is of the order of  $10^7-10^8$
el. cm$^{-3}$ \citep{Vanlandingham1996,Schwarz2007}. Since most atoms are
in the  ground term, assuming same emitting volumes, one  can  write:

\smallskip
\noindent

\begin{equation}
N(^{7}Be\,{\sc II})/N(Mg\,{\sc
II})=I_{3132}/I_{2880}\times(3132/2800)\times
(q_{12}^{2800}/q_{12}^{3132})
\end{equation}

\smallskip
\noindent
The ratio requires the knowledge of the collisional rate coefficients
 for \beii~ and \mgii:

\smallskip
\noindent

\begin{equation}
q_{12} = ( 8.63 \times 10^{-6} \times \overline{\Omega}_{12} \times
e^{-\Delta E/KT_e}) /(g_{12} \times   T_e^{1/2})
\end{equation}

\smallskip
\noindent
where $g_{12}$ is the statistical weight (2J+1) of the lower level of the
transition, $T_e$ is the electron  temperature,  e$^{-\Delta E/KT_e}$ is the Boltzmann factor   and
$\overline{\Omega}_{12}$
  is the thermally averaged (effective) collision strength, an atomic
parameter whose value is  not accurately determined and, generally, is in
the range 0.1-10.

The collision strength of the \beii~ resonance transition is not commonly
included in the astrophysical literature and is not clearly determined.
In the following, we  assume  $\overline{\Omega} (3132)  =  11 \pm 4$
as a weighted average of several estimates (see Appendix  A).
Substitution of this  value in the previous relation, with T$_e$ = 10000 K
(0.861 eV), and $\Delta$E=3.96 eV,   provides  $q_{12}$($\lambda$3132) =
4.36
($\pm$ 1.7) $\times$ 10$^{-9}$.

For  \mgii~ the uncertainty of the effective collision strength
(at about 10,000 K) is lower:  in  the following (see Appendix A)  we
assume
$\overline{\Omega}$ (2800) =  16 $\pm$ 2 which, after substitution in the
previous relations  with $T_e$=10000 K  (0.861 eV)  and $\Delta$E=4.43
eV,  provides $ q_{12}$($\lambda$2800) =  (4.35 $\pm$ 0.9) $\times$
10$^{-9}$.

Incidentally, it is  worth pointing out that the  reported under-abundance
of oxygen in V838 Her compared to other novae,  and the consequent
weakness of the oxygen lines, the most important coolants in a nebula,
could have as a consequence  an electron temperature higher than the value
$T_e = 10^4$ which is  usually assumed.

For the  April spectra  using   the data in column 4 of Table 3  and the above
mentioned  \texttt{numpy}
\citep{Numpy}  python package we
derived  a
($\sigma$)$^{-1}$ \  weighted
average  emission ratio $I(3132)/I(2800)$ $\sim$  0.80 $\pm$  0.26  (or
0.67 $\pm$ 0.134 in the case of a ($\sigma$)$^{-2}$  weight). From these
values and the previous relations we find   that  the relative
N(\beviiii)/N(\mgii)  number ratio is $\sim$ 0.90 $\pm$ 0.46, where the
final error originates from the  propagation of the error in  the emission
intensity ratio and in the effective collision strengths,
$\overline{\Omega}_{12}$, for the \mgii~ 2800 and \beii~ 3132 transitions.

The two ions have similar ionization potential and we can safely assume
that  N(\bevii)/N(Mg) $\approx$ N(\beviiii)/N(\mgii). Since  the
abundance of magnesium  in V838 Her is close to solar, from the
relative
Mg abundance in the Sun \citep{Grevesse1998} we find  that  the
abundance of \bevii~ in the ejecta, compared to hydrogen in the Sun  (by
number) is

\smallskip
\noindent
\begin{equation}\nonumber
N(^{7}Be)/N(H) = 3.4 (\pm 1.7)\times10^{-5}
\end{equation}
\smallskip
\noindent
For the correction by the decay of \bevii~, see Sect. 6.2.4.


\subsubsection{ The \beviiii~$\lambda$3132/H${\beta}$ ratio}

To the best of our knowledge,  there  is no published information  about
the emission line intensity  of  H$\beta$ in the  spectra of V838 Her
taken in April. There are, however, two  indirect sources of  information:

\begin{itemize}

\item

\citet{Schwarz2007}  reported  a ratio  \heii~$\lambda$1640 over
H$\beta$,    I$_{1640}$/I$_{4861}$  = 6.9 for April 22, 1991, but did not
provide the  intensity of  H$\beta$. Since an {\sl IUE} SWP  spectrum
of the same date is available, using this ratio one could  estimate  the
intensity of H$\beta$.  However, \citet{Schwarz2007} found
the ratio using a reddening correction $E(B-V)=0.60$ mag instead of $E(B-V)=0.40$ we derived.  In this latter case the correct  ratio becomes
about 3.3  and  from our direct measurement  of I$_{1640}$ = 51.0
$\times$
10$^{-12}$  erg cm$^{-2}$ s$^{-1}$, we estimate  that
 I$_{4861}$ =15.4  $\times$ 10$^{-12}$ erg cm$^{-2}$ s$^{-1}$.

\item
\citet{Williams1994a}  found  I$_{4861}$ /I$_{4686}$ $\sim$   0.5  in April
25. Since the expected  theoretical ratio I$_{1640}$/I$_{4686}$  is  $\sim$
 7.0 \citep{Dopita2003}  we find that I$_{4861}$ $\sim$  3.6  $ \times
10^{-12}$  [erg cm$^{-2}$ s$^{-1}$]. This is a lower limit because for
earlier
epochs (i.e. 22  April) one expects a higher I$_{4861}$/I$_{4686}$  ratio,
and so a value of  I$_{4861}$ close to 5.0 $\times$  10$^{-12}$  erg
cm$^{-2}$
s$^{-1}$  appears appropriate.
\end{itemize}

Based on the two above mentioned  estimates,  we assume  I$_{4861}$  $\sim$
(10.0  $\pm$ 5.0)   $\times$  10$^{-12}$  erg cm$^{-2}$ s$^{-1}$. This
provides
a  ratio  I$_{3132}$/I$_{4861}$   =  0.8 $\pm$ 0.56 for April 22.

\noindent
The ratio of the reddening corrected emission intensities  can be used to
estimate the number of atoms and the  relative abundances  adapting to our
case  the  relation \citep{Mihalszki1983,Osterbrock2006}:

\smallskip
\noindent

\begin{equation}
N(^{7}Be\,{\sc
II})/N(H^{+})=3132/4861\times\alpha_{eff}^{4861}/q_{12}^{3132}\times
I_{3132}/I_{4861}
\end{equation}

\smallskip
\noindent
For $T_e$ $\approx$ 10000 and   $\log N_e \approx 6-7$,
the effective recombination coefficient is $\alpha_{eff}^{4861} \approx  3.05\times10^{-14} cm^3 s^{-1}$,
with
small variations  about these values of T$_e$ and N$_e$  \citep{Osterbrock2006}.  Since $
q_{12}^{3132}$ =
4.36 ($\pm$ 1.74)$\times$10$^{-9}$ we find

\smallskip
\noindent
\begin{equation}\nonumber
N(^{7}Be{\sc II})/N(H+)  =  3.6 \sim(\pm 2.4 )\times 10^{-6}.
\end{equation}
\smallskip

The above reported  value  is probably a lower limit for the \bevii~
abundance
because  at this epoch  some \bevii~ is also in the
$^{7}Be^{++}$  ionization stage (\beviiiii~ spectrum), due to its
relatively  low  second ionization potential (18.21 eV).

The final quite large error derives from the propagation of the
individual uncertainties  of the line intensities, especially
H${\beta}$,  and from  that of  the effective
collision strength  $\overline{\Omega}_{12}$  for  the
\beviiii~~$\lambda$3132 transition.

\noindent
This apparent value for April 22 must be corrected for the decay
 of \bevii, see Sect. 6.2.4.

\subsubsection{ The \beviiii~$\lambda$3132/\heii$\lambda$1640  ratio}

Another  constrain to the \bevii~ abundance  can be found  using the
observed emission line ratio $I_{3132}$/$I_{1640}$   in
the spectra of April. The above relation, ''mutatis
mutandis'', becomes:

\smallskip
\noindent

\begin{equation}
N(^{7}Be\,{\sc II})/N(He\,{\sc
II})=I_{3132}/I_{1640}\times3132/1640\times\alpha_{eff}^{1640}/q_{12}^{3132}
\end{equation}

\smallskip
\noindent
that, after substitution of the numerical values  \citep{Osterbrock2006}  i.e.

\noindent
\begin{equation}\nonumber
\alpha_{eff}^{1640}=8.1\times 10^{-13},~~  q_{12}^{3132}=4.36(\pm
1.7)\times10^{-9}
\end{equation}
\noindent
and of the  average $I_{3132}$/$I_{1640}$  ratio = 0.20 $\pm 0.05$ gives:

\smallskip
\noindent
\begin{equation}\nonumber
N(^{7}Be{\sc II})/N(He{\sc II}) = 7.1(\pm 3.3) \times 10^{-5} .
\end{equation}
\smallskip
The lack of knowledge about the ionization fractions is quite a problem
because  helium is also \hei~   and \bevii~ could be twice-ionized; it is
worth noting that neutral helium  has a higher ionization potential than
singly-ionized \bevii~, and \heii~ is   strong.  Therefore, on the one hand
the \bevii/\he abundance could be higher if most \bevii~ were \beviiiii, on
the other hand  \bevii/\he~ could be lower if most of helium  where\
\hei, with a possible compensation between the two effects.

It appears, however, that helium  is mostly \heii~ as results  from the ratio
\heii(1640)/\hei(5876)=35 in April 22 \citep{Schwarz2007}. In this case
(helium mostly \heii), from the above reported  relation, and noting
that
\citet{Schwarz2007}  found that  in V838 Her helium  is overabundant
by a factor close to 1.4 compared to solar,   (that is,   \he/H   $\sim$
0.14), we derive  a lower  limit:

\smallskip
\noindent
\begin{equation}\nonumber
N(^{7}Be)/N(H)=1.0(\pm 0.45)\times 10 ^{-5}.
\end{equation}
\smallskip
\noindent
See Sect. 6.2.4 for the correction for the decay of \bevii.

\subsubsection{The average \bevii~ and \livii~   abundances}

The observed abundance of \beviiii~ must be corrected for the
radioactive decay  after its synthesis, assumed to take place entirely in
the early outburst phases. The correction for the  half-life  of 53.22 days
is almost negligible for the early absorption spectra taken a few days
after the outburst. Instead, it is, on average, by a factor $\approx$  1.30
for the abundances found in April from the emission line ratio
I$_{3132}$/I$_{2800}$ of \beviiii/\mgii,  and I$_{3132}$/I$_{1640}$ 
of \beviiii/\heii~ (from day 12 to day 29), and by a factor 1.46 for the
abundance from the ratio I$_{3132}$/I$_{4861}$   of \beviiii/HI for  April
22. The final values are reported in Table 5.

These four  values are quite consistent, in view of the various
uncertainties that may have affected their calculations,   and indicate  a
weighted  average  \bevii/H ratio  of approximately of $\approx 2.5
\times10^{-5}$  (by number),  i.e. of  $\approx 1.7 \times10^{-4}$  by
mass, (note that more weight was  given to the results from the direct
 \beviiii/\mgii~ ratio ).

This corresponds to an overproduction of \bevii~ by factors  of   about 8
and 40, respectively,  in comparison with the  theoretical models of
(massive) CO  and ONe  novae  of \citet{Jose1998} and \citet{Hernanz1998}
which indicate  an average mass  ratio \bevii/H  close to 2 $\times$
10$^{-5}$
and  4 $\times$  10$^{-6}$,  respectively.

Since \bevii\ all converts into \livii\, the corresponding \livii/H
ratio
($\approx  2.5 \times10^{-5}$,  by number) indicates  an overabundance
by about 4  dex over the  \livii\ meteoritic value (\livii/H =
1.86$ \times 10^{-9}$, \citep{Asplund2009,Lodders2009}), that is  an
 overproduction by about 1 dex over the models
of \citet{Jose2016} (cfr their fig 4.12).
These results  fairly
agree  with  the overabundance of 4.7 dex  relative to  the meteoritic
abundance  of \citet{Molaro2016} for V5668 Sgr.
2.

If the ejecta mass is $\sim$ 1.0 $\times$ 10$^{-5}$ $M_{\odot}$
\citep{Vanlandingham1996,Schwarz2007}  and the   \livii~/H  mass ratio is
about $ 1.7 \times10^{-4}$, assuming m(H)/m(tot) $\sim$
5.6$\times$10$^{-1}$
\citep{Schwarz2007}, then the total ejected mass of  \livii~   is of  about
9.5 $\times$10$^{-10}$  $M_{\odot}$.

\begin{table}
\caption{The average  \bevii/H~ (number)  derived from ratios of absorption
and emission lines.
The values are corrected for the decay of \bevii. \label{tab}}
\begin{center}
\begin{tabular}{lrrrlrrr}
\hline
\hline
Stage & Method & \bevii/H~ \\
\hline
\hline
Abs.& \beviiii(3132)/\mgii(2800) &  4.6 ($\pm$ 1.5)$\times$ 10
$^{-5}$\\
Em. & \beviiii(3132)/\mgii(2800) & 4.4 ($\pm$ 2.2)$\times$10
$^{-5}$ \\
Em. &  \beviiii(3132)/\heii(1640) & 1.3 ($\pm$ 0.6)$\times$ 10
$^{-5}$\\
Em. &  \beviiii(3132)/HI(4861)  & 0.5 ($\pm$ 0.4)$\times$
10 $^{-5}$\\
\hline
\end{tabular}
\end{center}
\end{table}

\section{Discussion}

The possibility of the  synthesis of \bevii~ and  of  its transport during
outburst to the outer nova layers in a process similar to the
\bevii-transport mechanism  in  red giants
\citep{Cameron1955,CameronFowler1971} represented a long debated and
controversial issue in the studies of nova nucleosynthesis.   See
\citet{Jose2007,Jose2016} for a comprehensive review  and  detailed
considerations on this subject.

It is commonly accepted  that  \bevii~  is produced by the $\alpha$-capture
reaction  $^3$He($\alpha,\gamma$)\bevii~ of the $pp$2-chain. This is  the
only exception  to the absence of $\alpha$-captures in the $pp$ chains
\citep{Jose2007}. If the unstable \bevii~, with a half-life of 53.22
days, survives destruction via the  proton capture reaction
\bevii($p,\gamma$)$^8Be$ that would complete the $pp$3 chain, the
reaction
of the pp2 chain  is followed by an inner shell electron capture on
\bevii~ with production of \livii.

The hydrodynamic simulations  with a full reaction network of
\citet{Hernanz1996}, and \citet{Jose1998} confirmed the feasibility of the
\bevii~  transport mechanism in nova outburst.
\citet{Hernanz2000,Jose2002,Hernanz2004} also found that CO novae are
favoured in comparison with  ONe novae  for \bevii~ synthesis because
their
faster rise-evolution
to $T \approx 10^8$ K in the early outburst phases, driven by their larger
amount of $^{12}C$
content,   favours \bevii~ survival:
\bevii~ destruction through the  deadly $p$-capture reactions  is
prevented due to to the efficient role played by the inverse
photo-disintegrations on $^8B$($\gamma$,p)\bevii
\citep{Jose2007,Isern2011,Jose2016}.

Needless to say, a  critical aspect in the studies of nova OB is the
treatment of hydrodynamic effects such as the  convective transport induced
by the large amount of energy released through the CNO cycle. Convection
with its extension throughout the envelope plays a fundamental role not
only in the transportation of the energy released in the ignition core, but
also in the mixing of material at the core envelope interface
\citep{Truran1998,Hernanz2000,Starrfield2016}.

In the specific case of the \bevii-transport mechanism, the observation of
\bevii~ and \livii requires that \bevii~ has to be transported by
convection  to low-temperature zones in a short timescale.  Therefore, the
observed abundance of \bevii~  (and \livii) in the ejecta is sensitive to
the rate at which it is transported to  the outer, cooler layers   prior to
its decay to \livii~ \citep[see Starrfield, Iliadis \& Hix in
][]{BodeEvansbook} and see also \citet{Isern2011}. Regrettably, the
efficiency of mixing by convection is a  critical parameter \citep[see e.g. ][]{Boffin1993}
\citet{Hernanz1996} computed the production of \bevii~  during nova
outbursts by means of a hydrodynamic code and derived an average mass fraction of \bevii~ in the shell of about 10$^{-6}$ -
10$^{-5}$ M$_{\odot}$ (see table 4 in \citealt{Hernanz1998})  and so  about
 1$\times$ 10$^{-10}$ - 1.2 $\times$ 10$^{-11}$ M$_{\odot}$ of \bevii~ were
ejected in a CO and a ONe nova respectively.

\citet{Wanajo1999} also found that CO novae may produce about 10 times more
\bevii~  than ONe   novae, with  \bevii~ abundances  in mass fraction of
up to 10$^{-6}$ for $T_{peak} \sim$ 3-4 times 10$^8$   K,  but they found
that for the same $T_{peak}$ the  models with lower WD mass  produce more
\bevii~ than higher ones. It should be noted, however,  that these models
were criticized  by  \citep{Downen2013}, as based on
outdated thermonuclear reaction rates.

In principle,  this complex  theoretical scenario could  be
tested with observational data by comparing the \bevii\ yields in CO and
ONe novae,  respectively. Regrettably,  in the case of V838 Her the
observational  constrains, i.e., the positive detection of \beviiii,
the extremely  fast  light curve character, the depletion of oxygen,  and
the  strong enrichment of  sulphur and aluminium  \citep{Schwarz2007},
indicate  a   contradictory  scenario for the nature of the WD and  the
nova
character:

\begin{itemize}

\item
The presence of a  massive  CO nova  (M$_1$  $\leq$ 1.15  $M_{\odot}$) is
required by  models \citep{Hernanz1998}  as the most probable  site for
\bevii~  synthesis, characterized by larger \bevii\  production, by a
factor close to 10, compared  to ONe novae. Models also  indicate that
fast novae such as V838 Her  require oxygen enhancement by  strong
mixing
with the CO interface. In this respect, the depletion of oxygen in the
ejecta of V838 Her is a  disturbing piece of  evidence.

\item
The presence of a massive ONe WD in V838 Her is indicated by the
following facts:
a. All spectroscopic  studies pointed out the presence of
exceptionally strong [\neiii] 3869-3968 and  [\nev]  3346-3428  emission
lines in late phases, see in particular  fig. 4  in  \citet{Williams1994a};

b. The extremely fast character of the light curve and the  high
expansion velocity  require a  massive WD of approx  1.35 $M_{\odot}$,
that is, a ONe WD   \citep{Matheson1993,Vanlandingham1996,Kato2009};

c. The observed overabundances  of sulphur and aluminium  by a factor of
about 30 compared to solar  require mixing with an underlying  massive
ONe (and not CO) WD, because CO WDs are devoid  of  these  overabundant
nuclei \citep{Jose2004,Hernanz2005}.
\end{itemize}

As noted in Sect. 2,  the WD mass, a fundamental parameter in all  the
above  mentioned   considerations,  is poorly determined from direct
radial
velocity data despite the orbital high inclination. Clearly a new direct
estimate of the mass would be  of paramount importance for a direct test of
the validity of the theoretical expectations.

It is worth  mentioning that the  depletion of oxygen and the significant
enrichment in heavier elements such as sulphur and aluminium  was
considered as an observational evidence that breakout from the CNO cycle
 may have occurred  in a  massive ONe  WD \citep{Schatz2004,Schwarz2007,Glasner2009}. However,  modeling of the
outburst of V838 Her \citep{Downen2013,Champagne2014} indicated a
massive (1.34-1.35 $M_{\odot}$) ONe WD, without  compelling evidence for
breakout ($T_{peak}$ in the TNR close to $3 \times 10^8$ K).

It should be also noted that the hydrodynamic models of\
\citet{Politano1995} and \citet{Jose1998}  predict a sharp  reduction of
oxygen and significant increase of sulphur  as the WD mass increases,
although,  as pointed out by \citet{Schwarz2007},  in the case of V838 Her
 both models fail to produce the observed
mass fraction of sulphur.

In conclusion, the presence of  large amounts of \bevii~  represents
an
additional peculiar aspect of
the properties of V838 Her which was considered  as
a "unique' object
\citet{Schwarz2007} even before  the  detection of \beviiii.

Probably, this uniqueness is  an
indication that nova models are not yet fully  predictive of the entire
variety of possibilities that
nature offers.

We note that  \bevii~ has been positively detected so far in slow novae
only, see \citet{Tajitsu2016,Molaro2016}. However,
\citet{Izzo2018} detected \bevii~ also in the fast  ONe V407 Lup.
Therefore, these last observations, summarized in Table \ref{tab5},
indicate  that synthesis of \bevii~ may
take place both in slow and in very fast novae, and  that ONe  novae are
also candidate sites.

\begin{table}
\caption{The   \bevii/H~ (number)  for the three Novae with narrow
absorption components.
The original values from \citet{Tajitsu2015,Tajitsu2016} are corrected here for the
decay of \bevii. References are: (1) \citet{Tajitsu2015}; (2)
\citet{Molaro2016},  (3) \citet{Izzo2018} and (4) \citet{Tajitsu2016}.  \label{tab5}}
\begin{center}
\begin{tabular}{llrrllrr}
\hline
\hline

\multicolumn{1}{c}{Nova} &
\multicolumn{1}{c}{type} &
\multicolumn{1}{c}{day} &
\multicolumn{1}{c}{comp} &
\multicolumn{1}{c}{ {\bevii/H~} } &
\multicolumn{1}{c}{Ref} \\
&&&&&\\
\hline
\hline

V339 Del & CO & 47 & -1103& 1.9 $\times$ 10 $^{-5}$  & 1\\

 &  &47  &-1268  &3.2 $\times$ 10 $^{-5}$   & 1 \\

V5668 Sgr & CO &  58 & -1175 & 1.7 $\times$ 10 $^{-4}$  & 2 \\
 &  & 82 &-1500 &1.3 $\times$ 10 $^{-4}$  & 2 \\
 V2944 Oph  & CO &  80 &  -645 & 1.6 $\times$ 10 $^{-5}$  & 4 \\
V407 Lup  & ONe&  8  &-2030 & 6.2  $\times$ 10 $^{-5}$ & 3 \\

\hline
\end{tabular}
\end{center}

\end{table}

\section{Conclusions }

\begin{itemize}

\item
By a detailed analysis  of archival{ \sl IUE} spectra of V838 Her
we  provided proofs that \bevii~  was produced  in the outburst of
V838 Her.  For the first time  the resonance line of \beviiii~  is also
detected   in emission;

\item
From  abundance analysis using both  absorption and emission lines
of
\beviiii~  we derive an
average   \bevii/H ratio  of approximately $\approx 2.5 \times10^{-5}$  (by
number) or of  $\approx 1.7 \times10^{-4}$ by mass.   This corresponds to
an overproduction by a factor about 8 and 40, respectively,  if compared to
models of massive  CO and the ONe novae \citep{Jose1998,Hernanz1998};

\item
Since \bevii\ all converts into \livii\, the corresponding \livii/H
ratio
($\approx  2.5 \times10^{-5}$,  by number) corresponds to an  overabundance
by about 4  dex over the  \livii\ meteoritic value  and   an
 overproduction by about 1 dex in comparison with  the values  in fig 4.12
of \citet{Jose2016}.

\item
If the mass of the ejected shell is $\sim$1.0$\times$10$^{-5}$ $M_{\odot}$
\citep{Vanlandingham1996,Schwarz2007} the total ejected mass  of
\livii~ is of  about 9.5 $\times 10^{-10}$  $M_{\odot}$.

\item

  Models favour  higher
synthesis of \bevii~ in CO novae
compared to ONe
novae,  although for ONe WD of 1.35  $M_{\odot}$
the overproduction factors    are similar.
\citep{Jose2016}.  Our results and those of V5668 Sgr   by
\citet{Molaro2016} and  and V407 Lup by \citet{Izzo2018} indicate that
indeed  ONe novae  produce less   \bevii~ than CO novae
(although in larger  amounts  than models).

\item

The present detection of \bevii\  in V838 Her,  the
previous detections of  \bevii\  (or  \livii)  in V339 Del
\citep{Tajitsu2015}, V2944 Oph \citep{Tajitsu2016}, V5668 Sgr
\citep{Molaro2016,Tajitsu2016} and \citep{Tajitsu2016},  V1369 Cen \citep{Izzo2015}
and  the
 detection of \bevii~ in V1369 Cen \citep{Izzo2015}  and  V407 Lup \citep{Izzo2018}
 confirm  that
\bevii\ can be synthesized  in a variety of novae, including  slow novae
and very fast novae,  and that ONe novae are also candidate places.

\item
 The presence of  large amounts of \bevii,   with the
depletion of oxygen and the overabundance of sulphur  (possible
indications  of breakout in a  massive ONe WD),   adds another
peculiar  aspect to the properties of  V838 Her  and confirms
its  characteristics as   a  ''unique'' object.

\item
\citet{Jose2007} concluded their comprehensive  review on
nucleosynthesis in CNe explosion  by stating:  ''unambiguous detection of
\livii~ has become a challenge: theoretical models suggest  a huge
overproduction of such isotope, particularly in CO novae, but detection
faces the likely superposition of lines corresponding to different
species''. In this respect, it is worth pointing out that  the detection of
the \beviiii~ resonance lines in recently observed  novae  and  in V838 Her,
 observed with {\sc IUE},   has opened
a new,  promising and
alternative  way to confirm the  effectiveness of the whole process of
\livii~ nucleosynthesis in novae.
\end{itemize}

\section*{Acknowledgments}
We gratefully acknowledge  Elena Mason for  helpful discussions on
 the interpretation of the data.   We also  thank  John
Danziger,  Maria Grazia  Franchini  and Carlo
Morossi for valuable comments and Gabriella Schiulaz for checking the English language.

\bibliographystyle{mnras}

\appendix

\section{The  collision strength of   \beii~ $\lambda$3132 and \mgii~
$\lambda$2800}

The effective collision strength $\overline{\Omega}$ of the \beii~
$\lambda$3132
resonance transition
is not commonly listed in astrophysical literature.

\citet{Osterbrock2006} and \citet{Dopita2003} provide
collision
strengths for  the $2s-2p$  ($^2S-^2P^o$)  transitions for   ions of
the \lii~ iso-electronic sequence  (i.e. \civ,  \nv~and \ovi)  but not for
\beii. Extrapolation to \beii~ of the tabulated values for the CNO ions
indicates $\overline{\Omega} \approx$ 10-12. Other estimates can be found
in
 literature both from various atomic models   and from  experimental
cross sections, resulting  in a range of values for $\overline{\Omega}$
for  \beii; an assessment of collision strengths for lithium-like ions  was
made by \citet{McWhirter1994}, using   $\overline{\Omega}$ values
based on the experimental cross section  by \citet{Taylor1980} and
\citet{Mitroy1988}. Extrapolation of these values to KT  $\sim$ 1 eV
indicates $\overline{\Omega}$ $\sim$  10. The Los Alamos theoretical
on-line
tables
\footnote{http://aphysics2.lanl.gov/tempweb/lanl/},
for $KT \sim 4 eV$, indicate $\overline{\Omega}$ $\approx$ 12, while the
Universal Fit Formula of  Cochrane $\&$ McWhirter \citep{McWhirter1994}
indicates $\overline{\Omega} \sim$14.  In this study  we   assume
$\overline{\Omega}(3132)=11\pm4$ as a  weighted average of these
estimates.

It is worth noting, however, that while  the comparison between laboratory
measurements and theoretical work on electron impact coefficients   agrees
 with theory for most ions of the Li iso-electronic sequence, there
is a long standing discrepancy for \beii~ (Dere et al, 1997).

\smallskip
     For  \mgii~ the uncertainties regarding the effective collision
strength are lower:  \citep{Osterbrock2006} (table 3.3)   give
16.9.   Similar values are found in \citet{Mendoza1981}, \citet{Dopita2003}
 and  \citet{Sigut1995}.
    The Los Alamos theoretical estimates   give $\overline{\Omega} \sim$
20, while \citet{Smith1993} give  values of about  10-20 for the  collision
strength close to  4-5 eV. In  this study   we assume
$\overline{\Omega}$(2800)=16$\pm$2   as a weighted average of these values.

\end{document}